\documentclass{PoS}

\usepackage{bbm}
\usepackage{bm}
\usepackage{graphicx}
\usepackage{epsfig}
\usepackage{subfigure}
\usepackage{latexsym}
\usepackage{amsmath}
\usepackage{amsfonts}
\usepackage{amssymb}
\newcommand{\be}{\begin{equation}}
\newcommand{\ee}{\end{equation}}
\newcommand{\bea}{\begin{eqnarray}} % only untightened
\newcommand{\eea}{\end{eqnarray}}

\newcommand{\bmp}{\noindent\begin{minipage}{16cm}}
\newcommand{\emp}{\end{minipage}\vskip 7mm} % 7mm untightened
\def\lsim{\mathrel{\raise.3ex\hbox{$<$\kern-.75em\lower1ex\hbox{$\sim$}}}}
\def\gsim{\mathrel{\raise.3ex\hbox{$>$\kern-.75em\lower1ex\hbox{$\sim$}}}}

\newcommand{\intron}[1]{}%{#1}
\title{Perturbative improvement of SU(2) gauge theory with two Wilson fermions in
the adjoint representation}

\ShortTitle{Perturbative improvement with adjoint Wilson fermions}

\author{\speaker{Tuomas Karavirta}, Kimmo Tuominen\\
	Department of Physics, P.O.Box 35 (YFL), \\
        FI-40014 University of Jyv\"askyl\"a, Finland, \\
        and \\
        Helsinki Institute of Physics, P.O.~Box 64, \\
        FI-00014 University of Helsinki, Finland\\
	E-mail: \email{tuomas.karavirta@jyu.fi},
        \email{kimmo.i.tuominen@jyu.fi}}

\author{Anne-Mari Mykk\"anen, Jarno Rantaharju and Kari Rummukainen\\
	Department of Physics and Helsinki Institute of Physics,\\
	P.O.Box 64, FI-00014 University of Helsinki, Finland\\
	Email: \email{anne-mari.mykkanen@helsinki.fi},
               \email{jarno.rantaharju@helsinki.fi},
               \email{kari.rummukainen@helsinki.fi}}

\abstract{%
  We present a perturbative calculation of the                                 improvement coefficients of SU(2) gauge theory with adjoint representation Wilson-clover fermions and using
Schrödinger functional boundary conditions.  The computation
of the boundary improvement terms is necessary for the full 
$O(a)$ improvement.  With two flavours of adjoint representation
fermions this theory is called Minimal Walking Technicolor
model.
} 

\FullConference{The XXVIII International Symposium on Lattice Field Theory, Lattice2010\\
		June 14-19, 2010\\
		Villasimius, Italy}

\begin{document}

\section{Introduction}

Recently, there has been a lot of interest in gauge theories which are conformal or near conformal. This is in particular due to their applications in beyond the Standard Model model building, most notably Technicolor and unparticle physics. In particular, for non-supersymmetric Yang-Mills theories with higher fermion representations it has been suggested \cite{Sannino:2004qp} that an ideal candidate for a walking technicolor theory would be the one with just two (techni)quark flavours in two-index symmetric representation of SU(2) or SU(3). The former of these two is called the Minimal Walking Technicolor (MWT) model, and is the subject of the study presented in this paper. Initial studies of both of these theories have been performed on the lattice already \cite{Catterall:2007yx,DelDebbio:2008zf,Catterall:2008qk,Shamir:2008pb,Hietanen:2008mr,Hietanen:2009az,Fodor:2008hn}.
For a related study of a QCD-like theory with fundamental representation fermions see ref.~\cite{Appelquist:2007hu}. 
 
The initial lattice studies of MWT using the Schr\"odinger functional method to measure the evolution of the coupling suggest that the  
theory has an infrared stable fixed point
\cite{Hietanen:2009az,Bursa:2009we}. However, the initial investigations with Wilson fermions used unimproved actions and are therefore subject to large ${\cal O}(a)$ discretization errors. 
These can be removed using Wilson-clover action with a suitably tuned
clover (Sheikholeslami-Wohlert) coefficient and introducing a set
of Schr\"odinger functional boundary counterterms, 
following the methods introduced in refs.~\cite{Luscher:1985wf,Luscher:1992an,Luscher:1993gh,Sint:1995ch,Luscher:1996vw}
for QCD.

We improve the theory in two distinct stages: first, the
Sheikholeslami-Wohlert coefficient is evaluated 
with non-perturbative methods.  This computation is 
described in ref.~\cite{clover}.  
In this paper we present the results of the perturbative computation of the coefficients of the boundary improvement terms.
The results of the physics simulations
will be presented elsewhere.

\section{Model and ${\mathcal{O}}(a)$ improvement}
We use the basic Wilson lattice action
\begin{eqnarray}
	S_{0}=S_G+S_F,
	\label{eq:aktion}
\end{eqnarray}
where the standard Wilson plaquette action is
\begin{equation}
S_G=\frac{\beta_L}{4}\sum_p{\textrm{tr}}(1-U(p)),
%  S_G = \beta_L \sum_{x;\mu<\nu} \left [1 - \frac12 \Tr P_{x;\mu\nu}\right]
%  =\frac{\beta_L}{4}a^4\sum_{x} F_{\mu\nu}(x)F^{\mu\nu}(x)
%  +{\cal{O}}(a^6),
\end{equation}
for the gauge part. The sum runs over all oriented plaquettes $p$, and $U(p)$ is the parallel transporter around the plaquette $p$ written in terms of the link matrices $U_\mu(x)$.  

The Wilson fermion action, $S_F$, for $N_f$ (degenerate) Dirac fermions in the fundamental or adjoint representation of the gauge group is
\begin{equation}
  S_{\textrm{F}} =a^4\sum_x\bar{\psi}(x)(D+m_{q,0}{\mathbbm{1}})\psi(x),
  % \sum_{N_f} 
  %\sum_{x,y} \bar\psi_{f,x} M_{xy} \psi_{f,y} 
    \label{sf}
\end{equation}
where the usual Wilson-Dirac operator is
\be
D=\frac{1}{2}(\gamma_\mu(\nabla_\mu^\ast+\nabla_\mu)-a\nabla^\ast_\mu\nabla_\mu),
\ee
involving the gauge covariant lattice derivatives $\nabla_\mu$ and $\nabla_\mu^\ast$ defined as
\bea
\nabla_\mu\psi(x) &=& \frac{1}{a}[\widetilde{U}_\mu(x)\psi(x+a\hat{\mu})-\psi(x)],\\
\nabla^\ast_\mu\psi(x) &=& \frac{1}{a}[\psi(x)-\widetilde{U}^{-1}_\mu(x-a\hat{\mu})\psi(x-a\hat{\mu})].
\eea
Here $\widetilde U$ is the parallel trasporter in the 
appropriate fermion representation.

It is well known that this action suffers from $\mathcal{O}(a)$ discretization errors. These errors can be removed by introducing the Sheikholeslami-Wohlert -term \cite{Sheikholeslami:1985ij} in the action  
\bea
S_{\rm{impr}} &=& S_0+\delta S_{\textrm{sw}},\\
\delta S_{\textrm{sw}} &=& a^5\sum_x c_{\textrm{sw}}\bar\psi(x)\frac{i}{4}\sigma_{\mu\nu} F_{\mu\nu}(x)\psi(x),\label{swterm}
\eea
and tuning the coefficient $c_{\textrm{sw}}$ so that the ${\mathcal{O}}(a)$ effects in on-shell quantities cancel. To the lowest order in perturbation theory $c_{\textrm{sw}}=1$ \cite{Wohlert:1987rf}. Here $\sigma_{\mu\nu}=i[\gamma_\mu,\gamma_\nu]/2$ and $F_{\mu\nu}(x)$ is the symmetrised lattice field strength tensor. 
 
In the Schr\"odinger functional scheme the fixed spatially constant boundary conditions at times $t=0$ and $t=T$ will introduce further $\mathcal{O}(a)$ errors. These errors can also be removed by introducing new terms to the action and fine tuning the corresponding coefficients so that the $\mathcal{O}(a)$ contributions cancel. Complete analysis of all necessary counterterms has been presented in \cite{Luscher:1996sc}. Here we only list the needed counterterms, which are
\bea
\delta S_V&=&\frac{i a^5 }{4}c_{sw}\sum_{x_0=a} ^{L-a}\sum_{\vec{x}}\bar{\psi}(x)\sigma_{\mu\nu}\hat{F}_{\mu\nu}(x)\psi(x),\\
\delta S_{G,b}&=&\frac{1}{2 g_0 ^2}(c_s-1)\sum_{p_s}{\rm{Tr}}[1-U(p_s)]+\frac{1}{g_0 ^2}(c_t-1)\sum_{p_t}{\rm{Tr}}[1-U(p_t)],\\
\delta S_{F,b}&=&a^4 (\tilde{c}_s-1)\sum_{\vec{x}}[\hat{O}_s(\vec{x})+\hat{O}'_s(\vec{x})]+a^4 (\tilde{c}_t-1)\sum_{\vec{x}}[\hat{O}_t(\vec{x})-\hat{O}'_t(\vec{x})].
\eea
Here we have introduced the operators
\bea
\hat{O}_s(\vec{x})&=&\frac{1}{2}\bar{\psi}(0,\vec{x})P_-\gamma_k(\nabla^* _k+\nabla_k)P_+\psi(0,\vec{x}),\\
\hat{O}'_s(\vec{x})&=&\frac{1}{2}\bar{\psi}(L,\vec{x})P_+\gamma_k(\nabla^* _k+\nabla_k)P_-\psi(L,\vec{x}),\\
\hat{O}_t(\vec{x})&=&\left\{\bar{\psi}(y)P_+\nabla^* _0\psi(y)+\bar{\psi}(y)\overleftarrow{\nabla}^* _0 P_-\psi(y)\right\}_{y=(a,\vec{x})},\\
\hat{O}'_t(\vec{x})&=&\left\{\bar{\psi}(y)P_-\nabla_0\psi(y)+\bar{\psi}(y)\overleftarrow{\nabla}_0 P_+\psi(y)\right\}_{y=(T-a,\vec{x})},
\eea
where the projection operators are $P_\pm=(1\pm \gamma_0)/2$.
By tuning the coefficients $c_{sw},c_s,c_t,\tilde{c}_s,\tilde{c}_t$ to their proper values we can remove all the $\mathcal{O}(a)$ errors from our action.

The coefficient $c_{sw}$ can be determined non-perturbatively and
its computation for SU(2) and adjoint fermions is described in ref.~\cite{clover}.
For the electric background field which we consider, the terms proportional to $c_s$ do not contribute. Also, we set the fermionic fields to zero on the boundaries, and then the counterterm proportional to $\tilde{c}_s$ vanishes.

\section{Perturbative analysis of the boundary improvement}
All boundary coefficients have a perturbative expansion of the form
\be
c_x=1+c_x ^{(1)} g_0 ^2+\mathcal{O}(g_0 ^4).
\ee
Here we will determine the coefficients $\tilde{c}_t$ and $c_t$ to one-loop order in perturbation theory.

\subsection{Coefficient $\tilde{c}^{(1)}_t$}
We follow the analysis performed in \cite{Luscher:1996vw} for the fundamental representation. The result of \cite{Luscher:1996vw} is
\be
\tilde{c}_t^{(1)}=-0.0135(1) C_F,
\ee 
and this generalises to other fermion representations simply by replacing the fundamental representation Casimir operator $C_F$ with Casimir operator $C_R$ of the representation $R$ under consideration. In the case of MWT we have fermions transforming in the adjoint representation of SU(2), hence $C_R=C_A=2$. The results for different gauge groups and fermion representations are shown in table~\ref{table:pert_impro}.

\subsection{Coefficient $c^{(1)}_t$}
The coefficient $c^{(1)}_t$ can be split into gauge and fermionic parts 
\be
c_t^{(1)}=c_t^{(1,0)}+c_t^{(1,1)}N_f.
\ee
The contribution $c_t^{(1,0)}$ is entirely due to gauge fields and has been evaluated 
%for fundamental representation fermions 
in \cite{Luscher:1992an} for SU(2) and in \cite{Luscher:1993gh} for SU(3).  The fermionic contribution $c_t^{(1,1)}$ to $c_t$ has been evaluated for fundamental fermions in \cite{Sint:1995ch} both for SU(2) and SU(3). We have extended these computations for SU(2) and SU(3) gauge theory with higher representation fermions and for SU(4) gauge theory with fundamental representation fermions. The results for the nonzero improvement coefficients are tabulated in table \ref{table:pert_impro}. The numbers beyond the fundamental representation are new, while those for the fundamental representation provide a good check on our computations. For the application to MWT, the relevant numbers are the ones on the second line of table~\ref{table:pert_impro}.

Our results are consistent with the generic formula
\be
c_t^{(1,1)} \approx 0.019141(2T(R)),
\label{conjecture}
\ee
where $T(R)$ is the normalization of the representation $R$, defined as ${\rm{Tr}}(T^a_R T^b_R)=T(R)\delta^{ab}$. For the details of the numerical method used to determine coefficient $c_t^{(1,1)}$, we refer to the original literature where the method was developed and applied first for the pure gauge theory case in \cite{Luscher:1992an}, and later for fundamental representation fermions in \cite{Luscher:1992an,Sint:1995ch}. 
\begin{table}[h!bt]
\center
\begin{tabular}{|c|c|c|c|c|}
\hline
$N_c$ & rep. & $c_t^{(1,0)}$ & $c_t^{(1,1)}$ &  $\tilde{c}_t^{(1)}$ \\
\hline
2 & ${\bf{2}}$ & $-0.0543(5)$ & $0.0192(2)$ & $-0.0101(3)$ \\
2 & ${\bf{3}}$ & $-0.0543(5)$ & $0.075(1)$ & $-0.0270(2)$ \\
3 & ${\bf{3}}$ & $-0.08900(5)$ & $0.0192(4)$ & $-0.0180(1)$ \\
3 & ${\bf{8}}$ & $-0.08900(5)$ & $0.113(1)$ & $-0.0405(3)$\\
3 & ${\bf{6}}$ & $-0.08900(5)$ & $0.0946(9)$ & $-0.0450(3)$\\
4 & ${\bf{4}}$ &               & $0.0192(5)$ & $-0.0253(2)$\\
\hline
\end{tabular}
\caption{The nonzero improvement coefficients for Schr\"odinger functional boundary conditions with electric background field for various gauge groups and fermion representations.}
\label{table:pert_impro}
\end{table}

\begin{figure}
\center
\includegraphics[scale=0.45]{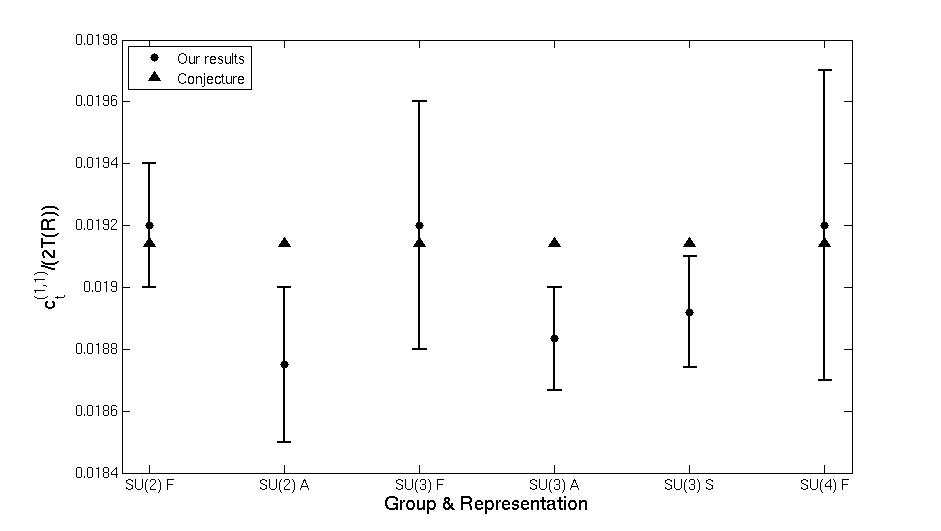}
\caption{Our results of $c_t ^{(1,1)}$ scaled with $2T(R)$ compared with conjectured value of $c_t ^{(1,1)}/(2T(R))$.}
\label{Pic}
\end{figure}

We have also plotted our results of $c_t ^{(1,1)}$ scaled with $1/(2T(R))$ against \eqref{conjecture} in figure \ref{Pic}. Although we were unable to achieve the accuracy of the original work \cite{Sint:1995ch}, our results are fully compatible for fundamental representation fermions. The figure also clearly indicates that $c^{(1,1)}_t$ scales with $2T(R)$. 

\section{Conclusions and outlook}

The search for conformal or near conformal gauge theories is phenomenologically motivated by the applications in model building beyond the Standard Model. Gauge theories with gauge groups SU(2) or SU(3) and with two Dirac fermions transforming under the two index symmetric representation are investigated by several lattice collaborations. When using Wilson fermions, the use of ${\cal O}(a)$ improved actions is desired. We have carried out this program for SU(2) gauge theory with two Wilson fermions in the adjoint representation, i.e. the MWT model. In this paper we presented the perturbative results for the improvement coefficients.

%The nonperturbative determination of the improvement coefficient %$c_{sw}$ and the results of the physics runs will be presented %elsewhere.

\acknowledgments
This work is supported by the Academy of Finland grant 1134018. T.K. is supported by the Magnus Ehrnrooth foundation and by University of Jyv\"askyl\"a Faculty of Mathematics and Science. A-M.M. is supported by the Magnus Ehrnrooth foundation and by Helsinki Institute of Physics. J.R. is supported by the V\"ais\"al\"a foundation.

\end{document}